\documentclass[12pt]{article}

\pdfoutput=1 %Will force arxiv to process with pdflatex

\usepackage{graphics, color}
\usepackage{graphicx}
\usepackage[colorlinks=true,urlcolor=blue,anchorcolor=blue,citecolor=blue,filecolor=blue,linkcolor=blue,menucolor=blue,linktocpage=true,pdfproducer=medialab,pdfa=true]{hyperref}
\usepackage[normalem]{ulem}
\usepackage{cite}

\usepackage{amsmath, amssymb, amsthm, amsbsy, bbm}

\theoremstyle{remark}

\theoremstyle{definition}

\newcommand{\Tr}{\mathrm{Tr}}

\newcommand{\AND}{\mathrm{\;\; \mathrm{and} \;\;}}

\newcommand{\scriplus}{\mathcal{I}^+}
\newcommand{\scrimin}{\mathcal{I}^-}
\newcommand{\Rscriplus}{\mathcal{I}^+_R}
\newcommand{\Rscrimin}{\mathcal{I}^-_R}
\newcommand{\Lscriplus}{\mathcal{I}^+_L}
\newcommand{\Lscrimin}{\mathcal{I}^-_L}
\newcommand{\scri}{\mathcal{I}}

\newcommand{\ket}[1]{| #1 \rangle}

\def\sl#1{\setbox0=\hbox{$#1$}           % set a box for #1 
   \dimen0=\wd0                                 % and get its size
   \setbox1=\hbox{/} \dimen1=\wd1               % get size of /
   \ifdim\dimen0>\dimen1                        % #1 is bigger
      \rlap{\hbox to \dimen0{\hfil/\hfil}}      % so center / in box
      #1                                        % and print #1
   \else                                        % / is bigger
      \rlap{\hbox to \dimen1{\hfil$#1$\hfil}}   % so center #1
      /                                         % and print /
   \fi}    
   
\newcommand{\fd}[3][]{\frac{\mathrm{d}^{#1} #2 \hfill}{\mathrm{d} {#3}^{#1} }}

\newcommand{\sfd}[3][]{{\mathrm{d}^{#1} #2 \hfill}/{\mathrm{d} {#3}^{#1} }}

\makeatletter
\renewcommand\d[1]{\mspace{2mu}\mathrm{d}#1\@ifnextchar\d{\mspace{-3mu}}{}\mspace{4mu}}
\makeatother

\newcommand{\bN}{\bar{N}}
\newcommand{\Fb}{F^{\mathrm{ATV}}}
\newcommand{\Mb}{M^{\mathrm{ATV}}_{\mathrm{Bondi}}}
\newcommand{\Ma}{M_{\mathrm{ADM}}}
\newcommand{\ym}{y^-}
\newcommand{\zm}{z^-}
\newcommand{\yms}{\ym_{\mathrm{sing}}}
\newcommand{\yp}{y^+}
\newcommand{\zp}{z^+}

\begin{document}

%------------------------------------------------------------------------%
%Title page

\begin{titlepage}
\bigskip
\bigskip\bigskip\bigskip
\centerline{\Large An Uneventful Horizon in Two Dimensions}
\bigskip\bigskip\bigskip
\bigskip\bigskip\bigskip

 \centerline{{\bf Ahmed Almheiri and James Sully}}
\bigskip\bigskip

%------------------------------------------------------------------------%
% ABSTRACT
\begin{abstract}
We investigate the possibility of firewalls in the Einstein-dilaton gravity model of CGHS. We use the results of the numerical simulation carried out by Ashtekar $et \ al.$ to demonstrate that firewalls are absent and the horizon is drama free. We show that the lack of a firewall is consistent because the model does not satisfy one of the postulates of black hole complementarity. In particular, we show that the Hawking radiation is not pure, and is completely entangled with a long-lived remnant beyond the last ray.
\end{abstract}
\end{titlepage}

\baselineskip = 16pt

\tableofcontents

\setcounter{footnote}{0}

\bigskip
\bigskip

%%%%%%%%%%%%%%%%%%%%%%%%%%%%%%%%%%%%%%%%%%%%%%%%%%%%%%%%%%%%%%%%%%%%%%%%%%%%%%%%%%%%%%%%%%%%%%%%%
%%%%%%%%%%%%%%%%%%%%%%%%%%%%%%%%%%%%%%%%%%%%%%%%%%%%%%%%%%%%%%%%%%%%%%%%%%%%%%%%%%%%%%%%%%%%%%%%%
\section{Introduction}
%%%%%%%%%%%%%%%%%%%%%%%%%%%%%%%%%%%%%%%%%%%%%%%%%%%%%%%%%%%%%%%%%%%%%%%%%%%%%%%%%%%%%%%%%%%%%%%%%
%%%%%%%%%%%%%%%%%%%%%%%%%%%%%%%%%%%%%%%%%%%%%%%%%%%%%%%%%%%%%%%%%%%%%%%%%%%%%%%%%%%%%%%%%%%%%%%%%

Almost forty years after Hawking's discovery that black holes radiate \cite{Hawking1975}, our understanding of the resulting black hole information paradox \cite{Hawking:1976ra} remains in a state of confusion. 
%The conflict between general relativity and quantum field theory has led some to consider of abandoning deep principles such as unitarity or locality [cite], and others to suggest the existence of remnants [cite]. There have been equally adamant rejections of these possibilities [cite]. The AdS/CFT duality has strongly motivated the conclusion that the solution to the paradox will maintain unitarity and will not involve remnants [cite].
Recent arguments by AMPS \cite{Almheiri} have given increased credence to the idea that the local semi-classical approximation breaks down at the horizon, resulting in a `firewall' and a failure of the equivalence principle. (Related arguments have also been made in \cite{Mathur2009,Giddings:2011ks, Giddings:2012bm, Giddings2012a,Braunstein2013}.) The AMPS firewall was motivated by the inconsistency of a set of long-held assumptions about the behaviour of black holes. These assumptions have been encapsulated as a set of postulates for black hole complementarity \cite{Susskind1993}. The particular postulates in conflict are: i) Unitarity of the S-matrix relating infalling matter to outgoing Hawking radiation ii) Validity of semi-classical field theory outside of the stretched horizon, and iii) Validity of semi-classical field theory in the infalling observer's local reference frame, i.e. no drama. While the authors of \cite{Almheiri} maintain that a firewall is the most conservative solution (as defended most recently in \cite{Almheiri2013}), much debate continues as to how the postulates might otherwise be modified to escape a contradiction \cite{Almheiri2013,Heemskerk:2012mn,Brown:2012un,Bousso:2012as,Nomura:2012sw,Mathur:2012jk,Chowdhury:2012vd,Susskind:2012rm,Bena:2012zi,Giveon:2012kp,Banks:2012nn,Ori:2012jx,Brustein:2012jn,Susskind:2012uw,Marolf:2012xe,Hossenfelder:2012mr,Nomura:2012cx,Hwang:2012nn,Avery:2012tf,Larjo:2012jt,Rama:2012fm,Page:2012zc,Papadodimas:2012aq,Nomura:2012ex,Giddings:2012gc,Neiman:2012fx,Saravani:2012is,Jacobson:2012gh,Susskind:2013tg,Harlow:2013tf,Page:2013dx,Kim:2013fv,Park:2013rm,Hsu:2013cw,Giddings:2013kcj,Kawai:2013mda,Avery:2013exa,Lee:2013vga,Brustein:2013xga,Iizuka:2013yla,Nomura:2013nya,Balasubramanian:2013rqa,Brustein:2013qma,Banks:2013cha,Mei:2013yta,Giveon:2013ica,Chowdhury:2013tza,Lowe:2013zxa,Verlinde:2013uja,Verlinde:2013vja,Maldacena:2013xja,Page:2013mqa,Shenker:2013pqa,Chapline:2013gza,Mathur:2013gua,Jensen:2013ora,VanRaamsdonk:2013sza}.

A simplified setting in which to attempt to better understand the information paradox and the possibility of firewalls is $1+1$-dimensional gravity. While one does not expect $1+1$-dimensional theories of gravity to be good models of black hole complementary in higher dimensions, the above postulates can be applied more broadly. If they all hold true in two-dimensions, then by the reasoning of \cite{Almheiri} they necessarily imply the existence of firewalls.

\begin{figure}
	\centering
		\includegraphics[height=8cm]{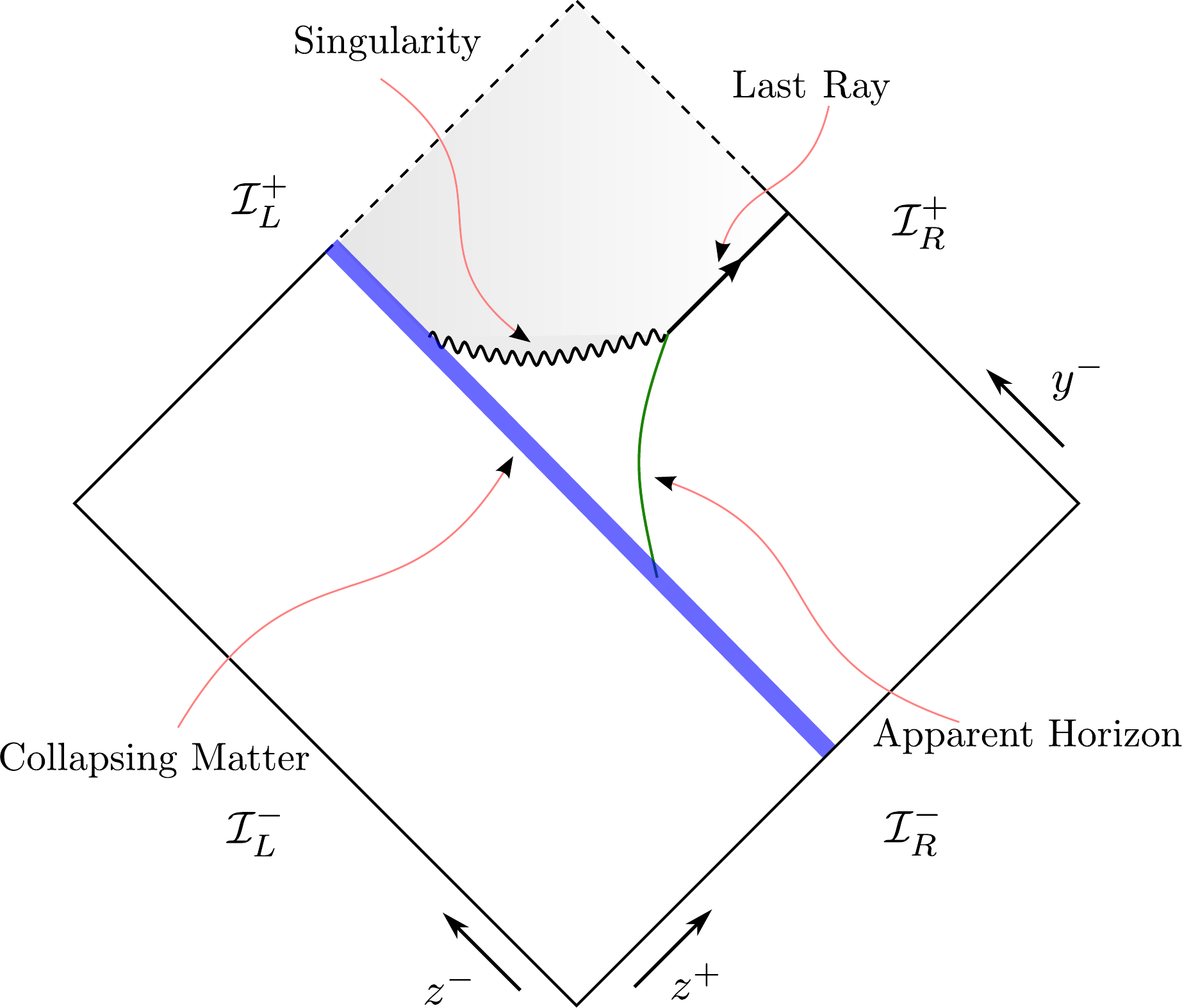}
	\caption{The geometry of an evaporating black hole in the mean field approximation of the CGHS model, as found by \cite{Ashtekar2008}. }
	\label{fig:CGHS}
\end{figure}

We choose, in particular, to study the CGHS model \cite{Callan1992} of dilaton-gravity with a large number of scalar matter fields in a background where a left-moving null shell of matter classically creates a black hole. This model (as well as other two-dimensional models) affords important simplifications not present in higher dimensions. Firstly, the metric and dilaton are not dynamical, but are determined in terms of the matter degrees of freedom. In fact, solutions to the classical equations of motion can be written in closed form. Secondly, the scalar field action is chosen such that the scalar fields only couple to the two-dimensional metric and thus decouple into left and right moving sectors that propagate freely. The asymptotic boundary also has more components than in higher dimensions, with both left and right portions of future null infinity $\scriplus$ and past null infinity $\scrimin$ (see Figure \ref{fig:CGHS}). This nicely separates the black hole information paradox into two separate questions \cite{Ashtekar2008}. The recovery of information sent into the black hole is a question about the unitarity of evolution of the state on $\Rscrimin$ to $\Lscriplus$. We will not address this question in this paper. Unitarity of Hawking radiation and the existence of firewalls is a question about the evolution of the state on $\Lscrimin$ to $\Rscriplus$. This is the question we will address here.\footnote{The corresponding cost of working in such a simplified model is a decreased explanatory power for understanding higher dimensions. We comment on this in the discussion.}

In the classical CGHS black hole solution, the last ray of the black hole singularity is at infinite affine parameter along $\Rscriplus$, but at finite parameter along $\Lscrimin$. Thus $\Rscriplus$ causally contains \emph{only} a proper subset of $\Lscrimin$ and we are left with a mixed state on $\Rscriplus$. However, the mean field analysis of \cite{Ashtekar2008,Ashtekar2011a,Ashtekar2011b} found that corrections to the Hawking radiation at late times brought the last ray of the singularity to a finite affine parameter along $\Rscriplus$. This suggests that, instead, $\Rscriplus$ could be unitarily equivalent to $\Lscrimin$. A complete understanding of the quantum evolution past the singularity would then give a pure state on $\Rscriplus$. Moreover, it was found that there is no singularity in the metric at the apparent horizon. The smooth horizon together with purity of the state on $\Rscriplus$ suggest a tension with the firewall argument.

In \cite{Ashtekar2008}, it was argued that the purity of the state on $\Rscriplus$ implies that there are \emph{not} remnants in the CGHS model. By this, they meant that the state on the entirety of $\Rscriplus$ is not entangled with another part of the future boundary. By contrast, we compute the entanglement entropy of an interval of $\Rscriplus$ containing the majority of the Hawking radiation and find it to be highly entangled with the remaining state to the future of the last ray of the black hole singularity, still on $\Rscriplus$. The remaining state beyond the last ray has a relatively small, universal (independent of the initial mass of the black hole) Bondi mass. This high-entropy, low-energy object is exactly a remnant. 

The large entanglement of the remnant evades the argument for firewalls because the Hawking radiation is not in itself pure. While we expect a unitary S-matrix that evolves the state at $\Lscrimin$ to $\Rscriplus$, the Hawking radiation is highly entangled with degrees of freedom behind the last ray of the black hole singularity. One is, of course, free include the remaining degrees of freedom in the definition of Hawking radiation, but what is important is the large entanglement remaining when the Bondi mass is small. This is better called a remnant.\footnote{We also emphasize---with the closely analogous case of a moving mirror----that the large entanglement of the Hawking radiation can be purified by a state beyond the last ray that locally looks like the vacuum \cite{Carlitz:1986nh,Holzhey1994}. There is not necessarily any further Hawking flux.} This conclusion is consistent with much early work on the CGHS model \cite{Callan1992,Giddings1992b,Banks:1992ba}.

Moving mirror models of Hawking radiation manifestly do not have firewalls, irrespective of measurement issues raised in \cite{Hotta:2013clt}, because the state is prepared precisely to be in the vacuum. Mirror trajectories that produce Hawking radiation always behave as some form of remnant \cite{Callan1994}.

Work in a very similar spirit to this paper has been carried out previously by \cite{Fiola:1994ir} in the RST model. Previous numerical studies of the CGHS model have been carried out by \cite{Piran:1993tq,Lowe:1992ed,Tada:1994wz,Hawking:1992ti}. Our work is able to draw stronger conclusions because of the numerical advances in \cite{Ashtekar2008,Ashtekar2011a,Ashtekar2011b} and their discovery of universal properties for suitably macroscopic black holes.  In recent work, \cite{Kim:2013fv}, it has been suggested that there could be a firewall outside the apparent horizon in related 2-dimensional models. We believe that there is sufficient control over the numerical simulations outside the apparent horizon such that this conclusion is not waranted. The arguments presented in this paper demonstrate instead how the firewall paradox is avoided.

An outline of the paper is as follows: in Section \ref{sec:CGHS}, we review the CGHS model and discuss the mean field theory results of \cite{Ashtekar2008,Ashtekar2011a,Ashtekar2011b}. In Section \ref{sec:entropy}, we review the entanglement entropy of an interval in a $1+1$-dimensional CFT and give a simple formula for the entanglement of the Hawking radiation at $\Rscriplus$ in the CGHS model. We also discuss the equivalence of related calculations for the entanglement entropy of radiation from a moving mirror. In Section \ref{sec:main}, we show that the entanglement entropy of the Hawking radiation in the CGHS model is large and scales like the ADM mass $\Ma$. We identify the modes in the interval that carry the excess entanglement, as well as the degrees of freedom that purify them across the last ray of the singularity. In Section \ref{sec:discussion}, we discuss the uplift of these solutions to higher dimensions and the connections between remnants in 2D and higher dimensions.

%%%%%%%%%%%%%%%%%%%%%%%%%%%%%%%%%%%%%%%%%%%%%%%%%%%%%%%%%%%%%%%%%%%%%%%%%%%%%%%%%%%%%%%%%%%%%%%%%
%%%%%%%%%%%%%%%%%%%%%%%%%%%%%%%%%%%%%%%%%%%%%%%%%%%%%%%%%%%%%%%%%%%%%%%%%%%%%%%%%%%%%%%%%%%%%%%%%
\section{A Review of the CGHS Model}\label{sec:CGHS}
%%%%%%%%%%%%%%%%%%%%%%%%%%%%%%%%%%%%%%%%%%%%%%%%%%%%%%%%%%%%%%%%%%%%%%%%%%%%%%%%%%%%%%%%%%%%%%%%%
%%%%%%%%%%%%%%%%%%%%%%%%%%%%%%%%%%%%%%%%%%%%%%%%%%%%%%%%%%%%%%%%%%%%%%%%%%%%%%%%%%%%%%%%%%%%%%%%%

In this section we review the 2 dimensional dilaton-gravity model of \cite{Callan1992}. The geometry of this model is specified by a  metric and a dilaton, given by $g_{a b}$ and $\phi$ respectively. This system closely resembles the one obtained from dimensional reduction of the s-wave sector of 4 dimensional gravity, but differs in the form of the dilaton potential. The benefit of studying such a model is that it is classically soluble in closed form, and is expected to provide qualitative insights into the s-wave sector of 4 dimensional gravity. The action of the geometric sector of this model is given by,
\begin{align}
S_{G} = \frac{1}{G} \int d^2x \sqrt{g} e^{- 2 \phi} \left(  R + 4 \nabla^a \phi \nabla_a \phi + 4 \kappa^2  \right) \, ,
\label{eq:gravityaction}
\end{align}
where $G$ is the two dimensional gravitational constant and $\kappa^2$ is a cosmological constant term. This action was originally obtained as the low energy effective action for string compactifications, where it describes the near horizon physics of extremal dilatonic black holes \cite{Giddings1992}.

Matter in this system is composed of $N$ scalar fields, $f^i$, whose action is, 
\begin{equation}
S_M = - \frac{1}{2} \int d^2 x \sqrt{g} \ \nabla^a f^i \nabla_a f^i \, .
\label{eq:matteraction}
\end{equation}
Note that the dilaton is absent in this action, and thus the scalar field can be viewed as living purely on the two dimensional spacetime given by the metric $g_{a b}$. This simplification ensures that the scalar field couples to the geometry only via the constraints.

%-------------------------------------------------------------------------------------------------%
\subsection{The Classical Solution}\label{subsec:CGHS_classical}
%-------------------------------------------------------------------------------------------------%

We will be adopting the conventions of Ashtekar $et \ al.$ used in \cite{Ashtekar2008,Ashtekar2011a,Ashtekar2011b}. We will be working in conformal gauge where the inverse metric takes the form $g^{a b} = \Omega \eta^{a b}$, and so all the metric information is encoded in $\Omega$. We will be considering the fields as living on a fiducial minkowski manifold $M^0$ with the flat metric $\eta_{a b}$ whose coordinates  are $z^+$ and $z^-$. This fiducial spacetime given by $(M^0, \eta)$ has null boundaries $\scri_{L,R}^{\pm}$, where the future(past) of $\scri_{L,R}^{-}$($\scri_{L,R}^{+}$) covers all of $M^0$. After redefining the dilaton as $\Phi \equiv e^{-2 \phi}$, and further defining $\Theta \equiv \Omega^{-1} \Phi$, the classical equations of motion take the form,
\begin{align}
&\partial_+ \partial_- \Phi + \kappa^2 \Theta = 0 \nonumber\\
&\Phi \partial_+ \partial_- \ln \Theta = 0 \nonumber\\
&\partial_+ \partial_- f^i = 0 
\end{align}
with the constraints,
\begin{align}
&-\partial^2_+ \Phi + \partial_+ \Phi \partial_+ \ln \Theta = \frac{G}{2} \sum_i (\partial_+ f^i)^2 \nonumber\\ 
&-\partial^2_- \Phi + \partial_- \Phi \partial_- \ln \Theta= \frac{G}{2} \sum_i  (\partial_- f^i)^2
\end{align}

Conformal gauge still leaves unfixed the conformal subgroup of diffeomorphisms. This is fixed by considering the solution,
\begin{align}
\Theta(z^{\pm}) &=  e^{\kappa (z^+ - z^-)}  \nonumber\\
\Phi(z^{\pm}) &= \Theta(z^{\pm}) - {G \over 2} \sum_i \int_{- \infty}^{z^+} d{z'}^+ e^{\kappa {z'}^+} \int_{-\infty}^{{z'}^+} d{z''}^+ e^{-\kappa {z''}^+} \ \left({\partial f^i_+ \over \partial {z''}^+}\right)^2
\end{align}
where $f^i_+$ is the left moving part of the solution $f^i = f^i_+(z^+) + f^i_-(z^-)$ of the scalar field equation. Setting $f^i = 0$ we see that this solution describes a flat metric with a linear dilaton; This is the so called throat limit of the extremal dilatonic black hole geometry in \cite{Giddings1992}.

As shown in \cite{Callan1992}, sending in a left moving scalar field shockwave creates a spacelike black hole singularity at the locus where the dilaton vanishes. Moreover, both $\scri_{L}^-$ and $\scri_{R}^+$ are complete with respect to $g_{a b}$, but the past of $\Rscriplus$ does not cover the entire spacetime, admitting a horizon for right moving modes. Thus classically, $\scri_{L}^{-}$ is not contained in the past of $\scri_{R}^+$. This is the basis of the information problem: the final state on $\scri_{L}^{+}$ is obtained from evolving the state on $\scri_{L}^{-}$ and tracing over part of the degrees of freedom that don't make it to $\scri_{R}^{+}$. Since we are working in two dimensions, left and right movers completely decouple and we can ask the question of unitarity for each one independently.

%-------------------------------------------------------------------------------------------------%
\subsection{The Mean Field Approximation}\label{subsec:CGHS_mean}
%-------------------------------------------------------------------------------------------------%

In \cite{Ashtekar2008,Ashtekar2011a,Ashtekar2011b}, Ashtekar $et$ $al.$ numerically studied the CGHS black hole solution at large $N$, including the back reaction of the Hawking radiation on the geometry. In the large $N$, mean-field approach they disregard quantum fluctuations of the geometry, but not that of matter. This approach was considered in earlier studies of both the CGHS model and close variants thereof \cite{Callan1992, Piran:1993tq, Russo:1992ht, Lowe:1992ed, Banks:1992ba}. The quantum state chosen for the scalar field is vacuum on $\Lscrimin$ and a coherent state with the classical profile $f_+^0$ on $\Rscrimin$, and thus $\langle \hat{f}^i_+ \rangle = f_+^0$. In keeping with matter fluctuations, the conformal anomaly \cite{Christensen:1977jc} in two dimensions results in a non traceless stress tensor which sources the equations of motion. The modified equations are,
\begin{align}
&\partial_+ \partial_- \Phi + \kappa^2 \Theta =  G \langle T_{+-} \rangle \equiv \bar{N} G \hbar \partial_+ \partial_- \ln \Phi \Theta^{-1} \label{mean:phipm} \\
&\Phi \partial_+ \partial_- \ln \Theta = -G \langle T_{+-} \rangle \equiv -  \bar{N} G \hbar \partial_+ \partial_- \ln \Phi \Theta^{-1} \label{mean:thetapm}
\end{align}
where $\bar{N} \equiv N/24$, and where the constraints,
\begin{align}
&-\partial^2_+ \Phi + \partial_+ \Phi \partial_+ \ln \Theta = 12 \bar{N} G (\partial_+ f^0_+)^2 \nonumber\\ 
&-\partial^2_- \Phi + \partial_- \Phi \partial_- \ln \Theta= 0
\end{align}
are specified on $\scri^-$. The shockwave is introduced as the coherent state coming in from $\scri_-^R$ given by
\begin{equation}
12 \bar{N} (\partial_{z^+} f^0_+)^2 = \Ma \delta(z^+)
\end{equation}
where $\Ma$ is the ADM mass of the resulting Black hole.

The mean field equations can be shown to provide a quantum corrected singularity \cite{Russo:1992ht, Banks:1992ba}. After a quick manipulation of  (\ref{mean:phipm}) and (\ref{mean:thetapm}), we obtain the equation 
\begin{align}
\kappa^2 \Theta - {\partial_+ \Phi \partial_- \Phi \over \Phi} &= (2 \bar{N} G \hbar  - \Phi) \partial_+ \partial_- \ln \Phi \Theta^{-1} \label{mean:correcteda}\\
 &\propto  (2 \bar{N} G \hbar  - \Phi) R_{g} \label{mean:correctedb}
\end{align}
where $R_{g}$ is the Ricci scalar of the metric $g_{a b} = \Phi^{-1} \Theta \eta_{a b}$. This implies a critical value for the dilaton, $\Phi_{cr} = 2 \bar{N} G \hbar$, where either the left hand side of (\ref{mean:correcteda}) vanishes or $R_g$ diverges. In the latter case we have a quantum singularity which occurs when the dilaton is nonzero. While we will assume that the full quantum evolution smoothly resolves this singularity, we believe our main conclusions will be unchanged should it be necessary to replace the singularity with a local boundary condition at $\Phi_{cr}$.\footnote{The boundary condition certainly will affect the evolution past the last ray, which we only speculate on, but it will not substantially affect the evolution before the last ray. This region is affected by a boundary condition in the linear dilaton region that is independent of the formation of the black hole. It is equivalent to a different choice of initial state.}

The size of an evaporating black hole can be tracked by the location of it's apparent horizon. Since the dilaton measures the area of the reduced 2-sphere, it can be used to define trapped points in 2 dimensions. Throughout the evaporation, the apparent horizon is then located at the future marginally trapped points, where $\partial_+ \Phi = 0$  and $\partial_- \Phi < 0$. Ashtekar $et$ $al.$ define the area on the future marginally trapped points as
\begin{equation}
a_{H} = {1 \over \kappa^2}\left(\Phi - 2 \bar{N} G \hbar \right)
\end{equation}
The shift proportional to $\hbar$ is induced by the mean field equations as a quantum correction to the singularity, which now occurs when $\Phi = 2 \bar{N} G \hbar$. This shift guarantees that $a_H$ shrinks to zero size at the singularity. 

Along $\Rscriplus$, the mean field theory equations imply a balance equation in terms of a quantum corrected Bondi mass and a quantum corrected Bondi flux \cite{Ashtekar2008}: 
\begin{equation}
\fd{\Mb}{\ym} = - \Fb
\label{eq:evap}
\end{equation}
where the corrected Bondi flux is given by
\begin{equation}
\Fb = \frac{\bN \hbar G}{2} \left[ \fd[2]{\ym}{\zm} \left( \fd{\ym}{\zm}\right)^{-2}\right]^2 
\label{eq:fatv}
\end{equation}
and the corrected Bondi mass is given by
\begin{equation}
\Mb =      \fd{B}{\ym} + \kappa B +  \bN \hbar G \fd[2]{\ym}{\zm} \left( \fd{\ym}{\zm}\right)^{-2} \, 
\label{eq:Mb}
\end{equation}
where $B$ is subleading term in the expansion of $\Phi$ at large $\yp$
\begin{equation}
\Phi =   e^{\kappa(\yp - \ym)} + B(\zm) + O(e^{-\kappa \yp})\, 
\label{eq:def_B}
\end{equation}
It is clear from the manifestly positive form of $\Fb$ that the Bondi mass uniformly decreases along $\Rscriplus$. While numerical computation showed \cite{Ashtekar2008} that the traditional Bondi mass, considered in \cite{Callan1992, Giddings:1994pj, Susskind:1992gd, Lowe:1992ed, Hayward:1992gi, Tada:1994wz, Hawking:1992ti}, can acquire a negative value at late times, the corrected mass was found to remain positive. 

The corrected Bondi flux is positive by definition, but this is not true of the traditional Bondi flux. The relation between the modified Bondi flux and the traditional one is given by
\begin{equation}
F^{\mathrm{Trad}} = \Fb  + \bar{N} \hbar G {\d \over \d y^-} \left[ { {\d{}}^2 y^- \over \d {z^-}^2}\left({\d y^- \over \d z^-} \right)^{-2}  \right]
\end{equation}

It was found \cite{Ashtekar2011b, Ori:2010nn} that the mean field theory equations admit a scaling symmetry that changes what is usually thought of as physically distinct parameters. This symmetry is  given by
\begin{align}
\tilde{\Theta}(z^+, z^-) &= \lambda \Theta(z^+, z^- + {\ln \lambda \over \kappa}) \nonumber\\
\tilde{\Phi}(z^+, z^-) &= \lambda \Phi(z^+, z^- + {\ln \lambda \over \kappa}) \nonumber\\
\tilde{N} &= \lambda N \nonumber\\
\tilde{f}^i_+(z^+) &= f^i_+(z^+)
\end{align}
where all the new fields satisfy the mean field theory equations. This induces the following change on the physical quantities
\begin{align}
&\Fb \rightarrow \lambda \Fb \nonumber\\
&\Ma \rightarrow \lambda \Ma \nonumber\\
&\Mb \rightarrow \lambda \Mb
\end{align}
This implies that the dynamics of the geometry depends only on the invariant quantities
\begin{align}
&F^* = \Fb /\bar{N} \nonumber\\
&M^* = \Ma /\bar{N} \nonumber\\
&M_{\mathrm{Bondi}}^* = \Mb /\bar{N} \nonumber\\
&m^* = M_{\mathrm{Bondi}}^*|_{\mathrm{Last \ ray}}
\end{align}

Whether a black hole is macroscopic or not depends  on how it's physical properties relate to the Planck scale. We adapt the conventions in \cite{Ashtekar2011b}. There is an ambiguity in defining the Planck mass and time in two dimensions, so we use the four dimensional definitions instead, which are $M^2_{Pl} = {\hbar/G_4}$ and $\tau^2_{Pl}  = G_4 \hbar$. From dimensional reduction, we have $G = G_4 \kappa^2$ and thus the Planck mass and time in two dimensional units are
\begin{equation}
M^2_{Pl} = {\hbar \kappa^2 \over G},  \ \ \mathrm{and} \ \ \ \tau^2_{Pl} = {G \hbar \over \kappa^2}
\end{equation}
An invariant relation to the Planck scale is to compare the time it takes a black hole of given ADM mass to evaporate via the non-backreacted Hawking flux, $F^{Haw} = \bar{N} \kappa^2 \hbar /2$, to the Planck time. Thus a black hole would be macroscopic if
\begin{equation}
M_{ADM}/F^{Haw} \gg \tau_{Pl} \ \rightarrow \ M^* \gg G \hbar M_{Pl}
\end{equation}

%-------------------------------------------------------------------------------------------------%
\subsection{Numerical Results}\label{subsec:CGHS_results}
%-------------------------------------------------------------------------------------------------%

The numerical simulation of \cite{Ashtekar2011b} bares a number of interesting results, some of which had been unanticipated. We focus on a few that are of most interest to the discussion at hand. Firstly, it was found that the dynamics was universal after a brief initial transient period. Physical quantities, invariants under the rescaling discussed above, would match a universal curve until the end of the evaporation process. It was found that for large enough $\Ma$, $\Mb$ approaches a universal value of $0.864\bar{N}$ in Planck units. This is a small mass, in that we expect it to evaporate in a time of order $\tau_{pl}$. Further, there is no `thunderbolt' curvature singularity along the last ray of the singularity \cite{Hawking:1992ti}: the Ricci scalar of the mean field theory metric is regular on the last ray and goes to zero as $\scri_R^+$ approached.

Further, in the mean field analysis the affine parameter along $\scri_R^+$ was found to be finite on the last ray, and thus $\scri_R^+$ is incomplete. This, along with the finiteness of the Ricci scalar on the last ray, imply that $\scri_R^+$ might very well be extendible past the last ray such that it is unitarily equivalent to $\Lscrimin$.

Upon close inspection of the Ricci scalar profile plots given in \cite{Ashtekar2011b}, it is clear that the Ricci scalar diverges only at the singularity and is finite everywhere else including the regions near the horizon. Thus firewalls are completely absent in this model. This at first seems to be at odds with the paradox of \cite{Almheiri}, which states that an infalling observer encounters high energy particles in the vicinity of the event horizon of a black hole. The rest of this paper is dedicated to showing that the dynamics of this model do not satisfy one of the postulates of black hole complementarity. The hawking radiation produced is in a mixed state and is in fact entangled with the region beyond the last ray. We describe this as a remnant scenario.

%%%%%%%%%%%%%%%%%%%%%%%%%%%%%%%%%%%%%%%%%%%%%%%%%%%%%%%%%%%%%%%%%%%%%%%%%%%%%%%%%%%%%%%%%%%%%%%%%
%%%%%%%%%%%%%%%%%%%%%%%%%%%%%%%%%%%%%%%%%%%%%%%%%%%%%%%%%%%%%%%%%%%%%%%%%%%%%%%%%%%%%%%%%%%%%%%%%
\section{Entanglement Entropy in 2D CFTs}\label{sec:entropy}
%%%%%%%%%%%%%%%%%%%%%%%%%%%%%%%%%%%%%%%%%%%%%%%%%%%%%%%%%%%%%%%%%%%%%%%%%%%%%%%%%%%%%%%%%%%%%%%%%
%%%%%%%%%%%%%%%%%%%%%%%%%%%%%%%%%%%%%%%%%%%%%%%%%%%%%%%%%%%%%%%%%%%%%%%%%%%%%%%%%%%%%%%%%%%%%%%%%

A very useful concept for understanding the state of the late-time Hawking radiation will be the entanglement entropy (or geometric entropy) of an interval \cite{Holzhey1993,Srednicki1993,Callan1994}. This entropy is simply the von Neumann entropy $S = -\Tr \rho_\Sigma \ln \rho_\Sigma$ of the density matrix $\rho_\Sigma$ for the state having traced out all degrees of freedom localized outside an interval $\Sigma$. 

For a $1+1$-dimensional CFT, it was shown \cite{Holzhey1993,Callan1994} that the entanglement entropy of an interval in the vacuum state is given by
\begin{equation}
S_\Sigma = \frac{c + \bar{c}}{6} \ln\left(\frac{\Sigma}{\sqrt{\epsilon_1 \epsilon_2}}\right)\, ,
\label{eq:interval}
\end{equation}
where $\epsilon_i$ are spatial UV cutoffs at either end. (An IR cutoff $\Lambda$ is also necessary, but here we work in the limit $\Sigma<<\Lambda$ where the entropy is independent of $\Lambda$.) 

This calculation of the entanglement entropy can easily be extended to non-vacuum states that are related to the vacuum by a conformal transformation \cite{Holzhey1994}. Because a conformal transformation is just a change of basis that respects the division of degrees of freedom into inside and outside the interval, the entanglement entropy will be invariant. Thus the entropy of $(\Sigma,\epsilon_1,\epsilon_2)$ in an appropriate non-vacuum state is given by
\begin{equation}
S_\Sigma = \frac{c + \bar{c}}{6} \ln\left(\frac{\widetilde{\Sigma}}{\sqrt{\widetilde{\epsilon_1} \widetilde{\epsilon_2}}}\right)\, ,
\label{eq:intervaltrans}
\end{equation}
where $(\widetilde{\Sigma}, \widetilde{\epsilon_1},\widetilde{\epsilon_2})$ are the transformed proper lengths of the interval and cutoffs in the conformally related vacuum. 

This entropy diverges as we remove the UV cutoff at either end. What we are really interested in is the renormalized entanglement entropy, which measures the excess entanglement relative to the vacuum:
\begin{equation}
S_{\mathrm{ren}} = S_{\mathrm{bare}} - S_{\mathrm{vac}} \, ,
\label{eq:ren1}
\end{equation}
which then is given by
\begin{equation}
S_{\Sigma,\mathrm{ren}} = \frac{c + \bar{c}}{6} \ln\left(\frac{\widetilde{\Sigma} \sqrt{\epsilon_1 \epsilon_2}}{\Sigma \sqrt{ \widetilde{\epsilon_1} \widetilde{\epsilon_2}}}\right)\, .
\label{eq:ren2}
\end{equation}

%-------------------------------------------------------------------------------------------------%
\subsection{Entanglement Entropy of Hawking Radiation in the CGHS model}\label{subsec:CGHS_entropy}
%-------------------------------------------------------------------------------------------------%

In the case at hand, we are interested in the entanglement entropy of the right moving modes across a region of $\Rscriplus$, given in the affine coordinate $\ym$ along $\Rscriplus$ as $\left[ \ym_1,\ym_2 \right]$. The interval is chosen to contain the vast majority of the Hawking flux. Let the conformal transformation that takes us to the vacuum be given by $\widetilde{y}^- = f(\ym)$ along $\Rscriplus$. To compute the entanglement entropy of this region, we consider some spacelike interval bounded by $\left[\ym_1,\ym_2\right]$ that is asymptotically close to $\Rscriplus$ so that
\begin{equation}
\frac{\widetilde{\Sigma}}{\Sigma} = \frac{f(\ym_2)-f(\ym_1)}{\ym_2 - \ym_1}\; , \AND \frac{\epsilon_1 \epsilon_2}{\widetilde{\epsilon_1} \widetilde{\epsilon_2}} = \frac{1}{f'(\ym_1)f'(\ym_2)} \, ,
\label{eq:trans}
\end{equation}
giving
\begin{equation}
S_{\Sigma,\mathrm{ren}} = \frac{c}{12} \ln\left(\frac{(f(\ym_2)-f(\ym_1))^2}{(\ym_2 - \ym_1)^2 f'(\ym_1)f'(\ym_2)} \right)\, .
\label{eq:ren3}
\end{equation}
The transformation that takes us to the vacuum at $\Rscriplus$ is exactly the transformation that makes the affine parameter along $\Rscriplus$ agree with that along $\Lscrimin$: $\zm = f(\ym)$. Moreover, for large negative $\ym$ there is no radiation emitted so that the affine parameters along $\Rscriplus$ and $\Lscrimin$ agree. Thus, if we take a long interval such that $\ym_1$ is sufficiently negative, then we can conclude
\begin{equation}
\frac{(f(\ym_2)-f(\ym_1))}{(\ym_2 - \ym_1)} \approx 1  \AND f'(\ym_1) \approx 1 \, ,
\label{eq:simplifcation}
\end{equation}
to arbitrarily high precision. We conclude that the renormalized entanglement entropy takes the simple form
\begin{equation}
S_{\Sigma,\mathrm{ren}} = - \frac{c}{12} \ln\left( f'(\ym_2) \right) = \frac{c}{12} \ln\left( \fd{\ym}{\zm} \right) \bigg\rvert_{\ym_2}\, .
\label{eq:entropyfinal}
\end{equation}

The independence of the renormalized entropy from moving $\ym_1$ further to the past is sensible. We expect the Hawking radiation to be largely independent of IR details of how the black hole was formed. Moreover, the invariance of the entropy as we extend the interval farther into the past implies that the excess entanglement is with degrees of freedom past the last ray, not at large negative $\ym$. Note, however, that there is a more complicated dependence of the renormalized entropy on \emph{large} relative changes in distances to the past and future boundaries for certain classes of remnants. We discuss this briefly in \ref{subsec:ent_modes}.

When $\ym$ is sufficiently negative, the entropy also depends entirely on the change in the cutoff on $\Rscriplus$ relative to $\Lscrimin$. Since we are considering the entropy of free right-moving fields, the entanglement of degrees of freedom inside and outside the interval cannot change from $\Rscriplus$ to $\Lscrimin$. In terms of modes on $\Lscrimin$, we find excess entanglement above the vacuum because, by fixing the proper length of the cutoff, we are now counting the entanglement of modes on $\Rscriplus$ that were above the cutoff on $\Lscrimin$ \cite{Fiola:1994ir}. However, when mapped forward to $\Rscriplus$, the identification of the modes carrying the excess entanglement is different. We elaborate on this point in \ref{subsec:ent_modes}.

%-------------------------------------------------------------------------------%
\subsection{Entanglement Entropy of radiation from Moving Mirrors}\label{subsec:mirrors}
%-------------------------------------------------------------------------------%

The above concepts are nicely illuminated by examining the closely analogous example of radiation from a moving mirror in $1+1$-dimensions \cite{Holzhey1994}. We describe here an equivalence between calculating the entanglement entropy in the CGHS model where the future and past affine parameters are related by $\ym = f(\zm)$ and doing the same computation for free scalar fields reflecting off a mirror with trajectory given by $\yp = f(\ym)$.\footnote{Classically there is also an equivalence of the flux produced by the two. With the corrected Bondi flux, this is no longer the case.} Most importantly, moving mirrors allow us to more tangibly discuss the generic types of behaviour we can expect after the last ray in the CGHS model.

Consider a mirror that moves in a flat background, $\d s^2 = - \d \ym \d \yp$, along the trajectory $\yp= f(\ym)$. The mirror provides perfectly reflecting boundary conditions for free scalar fields set in the vacuum along $\Rscrimin$. Again we are interested in the entropy of a spacelike interval spanning the null interval $\left[\ym_1,\ym_2 \right]$. The entropy can be calculated again by using a conformal transformation, $\widetilde{y}^- = f(\ym)$, to map the configuration to the vacuum where the mirror trajectory is given by $\yp = \widetilde{y}^-$. As in the CGHS model considered above, the entropy relative to the vacuum is determined by the rescaling of the interval and cutoffs under the conformal transformation. Exactly as before, we find
\begin{equation}
S_{\Sigma,\mathrm{ren}} = \frac{c}{12} \ln\left(\frac{(f(\ym_2)-f(\ym_1))^2}{(\ym_2 - \ym_1)^2 f'(\ym_1)f'(\ym_2)} \right)\, .
\label{eq:ren_mirror}
\end{equation}
We are interested in a mirror trajectory such that for some $Y_1^-$, we have $f(\ym) = \ym$ for all $\ym < Y_1^-$. Then given an interval $\left[\ym_1,\ym_2 \right]$ such that $\ym_1 << Y^-_1$, the entropy again reduces as in the CGHS model
\begin{equation}
S_{\Sigma,\mathrm{ren}} = - \frac{c}{12} \ln\left( f'(\ym_2) \right) \, .
\label{eq:entropy_mirror_final}
\end{equation}
Now suppose further that our mirror accelerates from rest until reaching a constant left-moving velocity, ie. $f'(\ym)<1$. It has a greater entanglement entropy than the vacuum. As this entropy is unchanged when we take $\ym \rightarrow - \infty$, the excess entanglement is with modes across the boundary at $\ym_2$. There is an interval of radiation along $\Rscriplus$ when the mirror is accelerating followed by what locally appears to be the vacuum. The large amount of entanglement between the radiating interval and the vacuum interval indicates that we should understand this trajectory as the analog of a remnant \cite{Holzhey1994}.

\begin{figure}
	\centering
		\includegraphics[height=8cm]{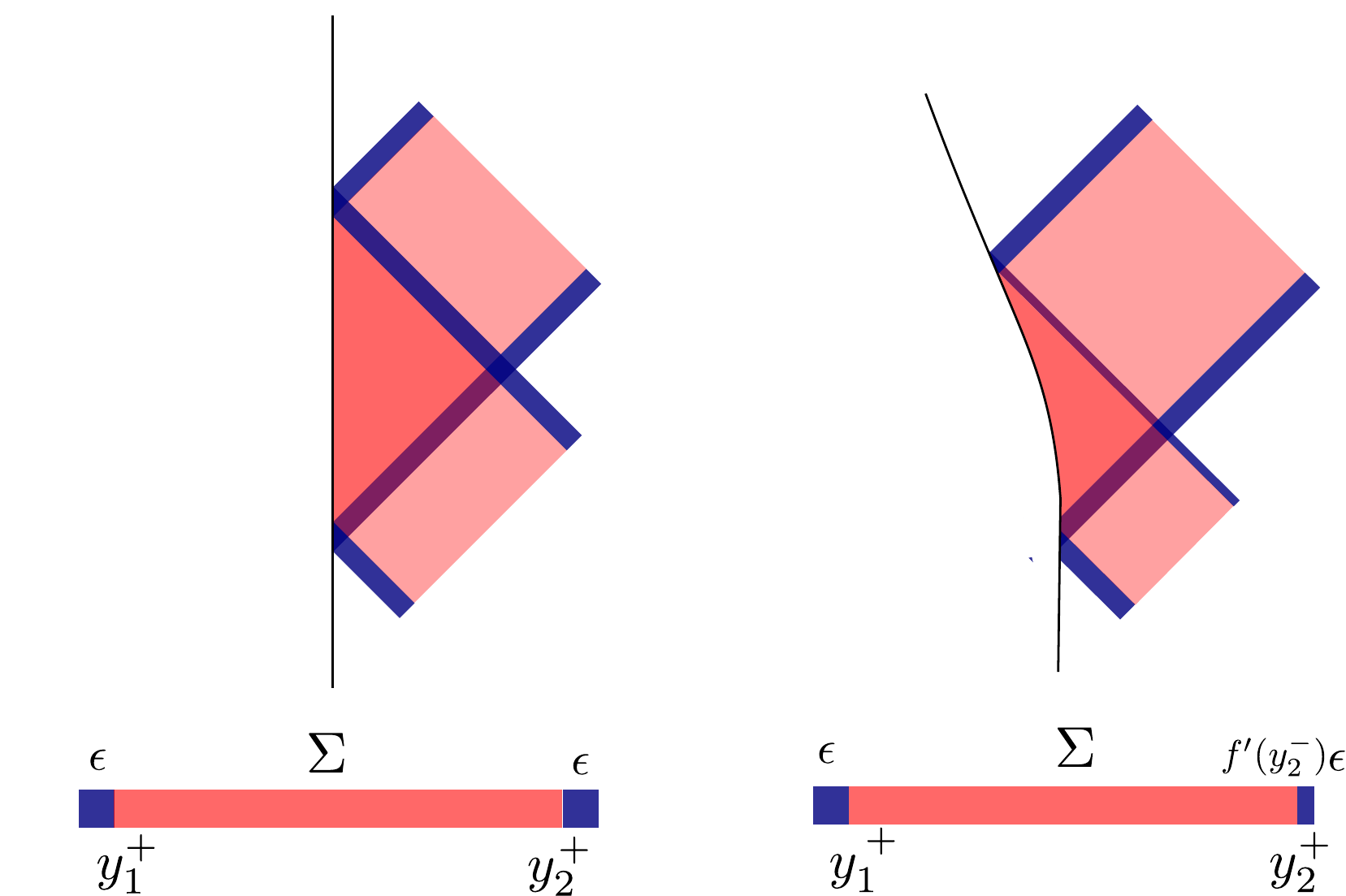}
	\caption{A set of right-moving modes on an interval of $\Rscriplus$ can be mapped back to a set of left-moving modes in the vacuum on $\Rscrimin$. The same interval on $\Rscriplus$, with different mirror configurations, maps back to different length intervals on $\Rscrimin$. In this specific example, we see the intervals on $\Rscrimin$ differ, at leading order,  only by the length of the cutoff at the future bondary.}
	\label{fig:mirror2}
\end{figure}

As in the CGHS model, to better understand where the excess entanglement comes from, we can simply map all questions of entanglement of different states on $\Rscriplus$ to the entanglement of different intervals on $\Rscrimin$. We consider a fixed interval on $\Rscriplus$ for our moving mirror and for the non-moving mirror vacuum. The two cases are illustrated in Figure \ref{fig:mirror2}. Because the majority of the length of the interval is contained in the region where the mirror is not moving, the length of the interval on $\Rscrimin$ is the same at leading order for both the vaccuum and the moving mirror. However, the cutoff at $\ym_2$ is rescaled by a factor of $f'(\ym_2)$, relative to the vacuum, when traced back to $\Rscrimin$. The excess entropy for the moving mirror can thus be understood, in terms of modes on $\Rscrimin$, to be modes that have been pulled down from above the UV cutoff.

Note that this example displays interesting behaviour for the renormalized entropy when we make large changes in the location of boundaries. As we increase $\ym_2$, at leading order there is no change in $S_{\Sigma,\mathrm{ren}}$. But, as we continue to move $\ym_2$ farther along the region of constant $f'(\ym)$, eventually the length of portion of our interval in this region becomes comparable to that in the region where the mirror is not moving, and our approximation breaks down. In this regime, $S_{\Sigma,\mathrm{ren}}$ is shrinking. If we were to further increase $\ym_2$, we eventually reach a regime where
\begin{equation}
\frac{\widetilde{\Sigma}}{\Sigma} \approx f'(\ym_2)\; , \AND \frac{\epsilon_1 \epsilon_2}{\widetilde{\epsilon_1} \widetilde{\epsilon_2}} \approx \frac{1}{f'(\ym_2)} \, ,
\label{eq:trans2}
\end{equation}
so that 
\begin{equation}
S_{\Sigma,\mathrm{ren}} \approx  \frac{c}{12} \ln\left({{f'(\ym_2)}}\right) <  0 \, .
\label{eq:entropy_mirror_2}
\end{equation}
The interval is now actually even less entangled than it would be in the vacuum state. The entropy can be completely rejuvenated now by simply moving $\ym_1$ sufficiently further to the past. While this isn't surprising when considering the modes on $\Rscrimin$, it is less intuitive in terms of the state on $\Rscriplus$. We explain this long-range behaviour in section \ref{subsec:ent_modes}.

As a separate example, we can also consider a mirror trajectory that, after accelerating to a constant velocity and producing Hawking radiation, quickly decelerates back to zero velocity in a short affine interval. We take $\ym_2$ after this deceleration, and note that length of the corresponding interval on $\Rscrimin$ is the same as in the vaccuum at leading order. Likewise, the cutoffs are also now the same as in vacuum. Thus $S_{\Sigma,\mathrm{ren}}\approx 0$. In this case the excess entanglement in the accelerating region is clearly purified by degrees of freedom that live in a short decelerating interval. This example does not exhibit the interesting long-range entanglement as did the previous case.

%%%%%%%%%%%%%%%%%%%%%%%%%%%%%%%%%%%%%%%%%%%%
%%%%%%%%%%%%%%%%%%%%%%%%%%%%%%%%%%%%%%%%%%%%
\section{CGHS Remnants}\label{sec:main}
%%%%%%%%%%%%%%%%%%%%%%%%%%%%%%%%%%%%%%%%%%%%
%%%%%%%%%%%%%%%%%%%%%%%%%%%%%%%%%%%%%%%%%%%%

We now show that the renormalized entanglement entropy of the black hole radiation on $\Rscriplus$ is large and scales like $M_{ADM}$. This excess entanglement (above the vacuum) is between the radiation and the state in the causal future of the black hole. This remaining object has a large entropy and a small Bondi mass; it is a remnant.

We can rewrite the mean-field corrected Bondi flux as
\begin{equation}
\Fb = \frac{\bN \hbar G}{2} \left[ \fd{\ln(\fd{\ym}{\zm})}{\ym}\right]^2 = \frac{\bN \hbar G}{2} \left[ \fd{s(\ym) }{\ym}\right]^2 \, .
\end{equation}
where we have defined the function $s(\ym) =  \ln(\fd{\ym}{\zm})$.

When $M^*_{\mathrm{Bondi}} >> M_{Pl}$, we expect that the semi-classical description remains approximately valid and the black hole radiates at a constant rate set by the temperature $T = \kappa \hbar$. This is borne out by the numerical simulations \cite{Ashtekar2008}, where it is found that, while $M^*_{\mathrm{Bondi}}>>M_{Pl}$, and after the formation of the apparent horizon, then $\sfd{\ym}{\zm}\approx \left[\kappa(z_{\mathrm{sing}}^- - z^-)\right]^{-1}$ so that $\sfd{s}{\ym} \approx \kappa$.

To leading order, for a large black hole, we can then write \eqref{eq:evap} as
\begin{equation}
\fd{\Mb}{\ym} \approx - \frac{\kappa \bN \hbar G}{2} \fd{s }{y^-} \, .
\label{eq:evaps}
\end{equation} 
Integrating this equation from $\ym = - \infty$ to the last ray gives
\begin{equation}
\frac{2}{\kappa \bN \hbar G} \left( \Ma - \Mb|_{\mathrm{Last\, ray}} \right) \approx s(\ym_{\mathrm{sing}})- s(-\infty) \, .
\label{eq:integral}
\end{equation}
Because the future and past affine parameters agree at large negative $\ym$, and recalling that \cite{Ashtekar2008} found that $\Mb|_{\mathrm{Last\, ray}}$ was universal and small as compared to our choice of $\Ma$, ($\Mb|_{\mathrm{Last\, ray}} \approx 0.84 \bN M_p$), this gives at leading order
\begin{equation}
S = \frac{N}{12} s(\ym_{\mathrm{sing}}) \approx \frac{4 \Ma}{\kappa \hbar G}   \, .
\label{eq:result}
\end{equation}
We immediately recognize the left hand side of this equation as the renormalized entanglement entropy of a large segment of $\Rscriplus$ ending at $\yms$. It contains contains almost all of the black hole radiation. The entropy scales with $\Ma$ and so is large in comparison to $\bN M_p$. A similar result was previously argued for using thermodynamic arguments (and assuming necessary corrections to the late time Hawking flux) \cite{Giddings:1994pj}.

Note that from \cite{Ashtekar2008}, we can extract this result from their numerical computation of $\sfd{\ym}{\zm}$ for the infinite family of black holes with $\Ma=8 \bN M_p$. One finds that
\begin{equation}
S = 13.4 \frac{N}{12}  \, .
\label{eq:}
\end{equation}
This agrees well with our approximate analytic calculation, which in this case gives $S = \frac{4}{3} N$ (the numerical simulation uses units where $\kappa = \hbar = G =1$). The fact that our analytic result is slightly larger is expected as our approximation underestimates the rate of flux at late times as compared to the numerical simulation.
 
As was argued earlier in Section \ref{sec:entropy}, this entropy is insensitive to our choice for the boundary of the interval at some large negative $\ym$. Thus, it is clear that the excess entanglement is with the state to the future along $\Rscriplus$. Moreover, the corrected Bondi mass is small at $\yms$ and is always decreasing. We identify such a small mass state that has a large entanglement entropy as a remnant. While the separation of degrees of freedom on $\Rscriplus$ into Hawking radiation and a remnant is somewhat artificial, the salient point is that the entropy is increasing as we include more of the Hawking radiation in our interval and that the entropy is still large when the remaining Bondi mass is small. 

It's not possible to describe what the degrees of freedom that are entangled with the Hawking radiation look like on $\Rscriplus$ without knowledge of $\sfd{\ym}{\zm}$. Two general possibilities were described in the analogous example of the moving mirror, where it either continued at constant velocity, or decelerated to zero velocity. Indeed, if we had a positive mass theorem for the modified bondi mass $\Mb$, then the modified flux would necessarily have to approach zero. This is equivalent to the requirement of $\sfd{\ym}{\zm} = k$. 

Still assuming a positive mass theorem, we can go somewhat further: if the mirror is to return to zero velocity in the original frame, then it must happen slowly. Thus, the remnant is necessarily large on $\Rscriplus$. To see this, consider the integrated flux after the last ray
\begin{align}
\int_{\ym_{\mathrm{sing}}}^{\infty} F \d{\ym} &= \frac{\kappa \bN \hbar G}{2} \int_{\ym_{\mathrm{sing}}}^{\infty} \left( \fd{s}{\ym} \right)^2 \d{\ym} \nonumber \\
                                              &= \frac{\kappa \bN \hbar G}{2} \int_{s(\ym_{\mathrm{sing}})}^0 \left( \fd{\ym}{s}\right)^{-1} \d{s}
\label{eq:fluxagain}
\end{align}
A positive mass theorem would give the constraint
\begin{equation}
 \Mb|_{\mathrm{Last\, ray}} > \int_{\ym_{\mathrm{sing}}}^{\infty} F \d{\ym}
\label{eq:positivemass}
\end{equation}
and, as $s(\ym_{\mathrm{sing}}) \approx \frac{2}{\kappa \bN \hbar G}  \Ma$, we then must have that
\begin{equation}
\Delta \ym = \int_{s(\ym_{\mathrm{sing}})}^0 \left(\fd{\ym}{s} \right) \d{s} > \frac{\kappa \bN \hbar G}{2} \frac{s(\ym_{\mathrm{sing}})^2}{\Mb|_{\mathrm{Last\, ray}}} = s(\ym_{\mathrm{sing}}) \frac{\Ma}{\Mb|_{\mathrm{Last\, ray}}}
\label{eq:deltay}
\end{equation}
As both $s(\ym)$ at the last ray and the ratio ${\Ma}/{\Mb|_{\mathrm{Last\, ray}}}$ are large, we must have that the affine parameter over which the mirror returns to rest is large. We thus rule out the class of mirror trajectories where the mirror quickly returns to rest in the original frame.

Nevertheless, it is worth noting that despite the fact that the remnant is large in terms of the affine parameter $\ym$ on $\Rscriplus$, it can still occupy a much shorter interval in terms of $\zm$, because at the last ray we have $\sfd{\ym}{\zm}$ is exponentially large ($\sfd{\ym}{\zm} \approx e^{M^*}$).

%---------------------------------------------------------------%
\subsection{Identifying Entangled Modes}\label{subsec:ent_modes}
%---------------------------------------------------------------%

\begin{figure}
	\centering
		\includegraphics[height=9cm]{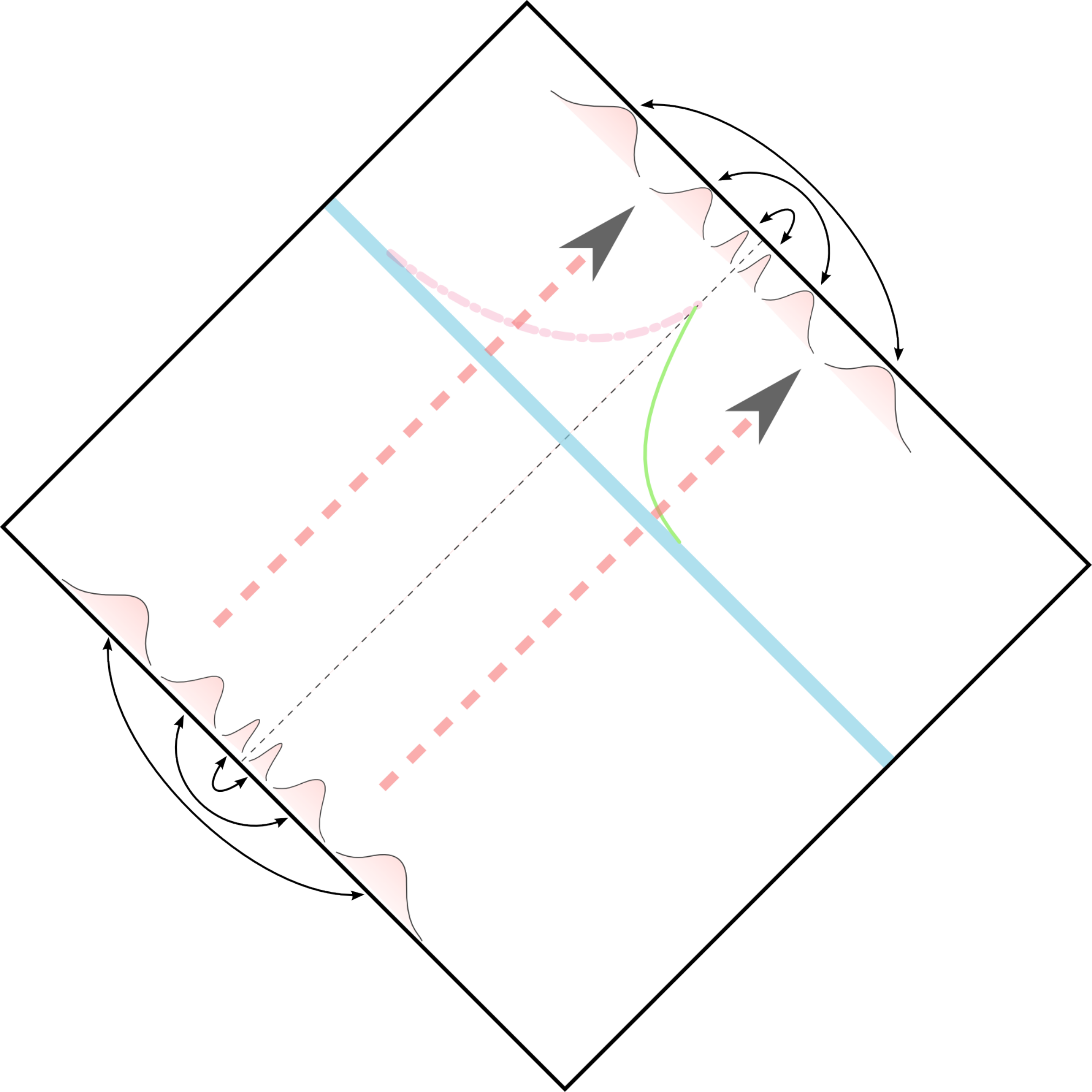}
	\caption{The radiated modes in the Hawking-like region are exactly the localized modes described in \cite{Giddings1992b}. They are entangled with partner modes across the last ray of black hole singularity.}
	\label{fig:EntModes}
\end{figure}

We wish to understand what the degrees of freedom are in the remnant region that purify the excess entanglement contained in the Hawking radiation. Recall that in the semi-classical treatment of the CGHS model \cite{Giddings1992b}, we can divide $\Lscrimin$ at the last ray into two semi-infinite intervals. We take the last ray here to be at $\zm=0$. As the last ray is at infinite affine parameter along $\Rscriplus$, a basis of modes to the past of the last ray is given by (letting $\kappa =1$)
\begin{equation}
v_\omega (\zm) = \frac{1}{\sqrt{2 \omega}} e^{i \omega \ln(- \zm)} \Theta(-\zm)
\label{eq:hawkingmodes}
\end{equation}
which are positive frequency for $\zm$ near $0$ (the region of Hawking radiation). A corresponding choice of modes can be made to the future of the last ray
\begin{equation}
\widehat{v}_\omega(\zm) = \frac{1}{\sqrt{2 \omega}} e^{-i \omega \ln( \zm)} \Theta(\zm)
\label{eq:bhmodes}
\end{equation}
We can also construct localized wavepackets
\begin{equation}
v_{jn}  = \epsilon^{-1/2} \int_{j\epsilon}^{(j+1)\epsilon} \d{\omega} e^{2 \pi i \omega n/\epsilon } v_\omega
\label{eq:localmodes}
\end{equation}
peaked at $\zm = - \exp(-2 \pi n/\epsilon)$ with width $\epsilon^{-1}$ in $\ln(-\zm)$ and frequency $\omega_j = j \epsilon$. One then finds \cite{Giddings1992b} that in this basis, for small $\epsilon$, the in-vacuum has the thermal form
\begin{equation}
\ket{0}_{\mathrm{in}} = Z^{-1/2} \sum_{n_{jn}} \exp\left(-\pi\sum_{jn}n_{jn} \omega_{j} \right) \ket{\widehat{n_{jn}}}\ket{n_{jn}}
\label{eq:thermalvac}
\end{equation}
where $n_{jn}$ are occupation numbers for the localized wavepackets.

Now in the mean field approximation, we can consider such wavepackets that are localized in the predominant region where the radiation is Hawking-like. These modes necessarily have the same entanglement as above and are purified by their partner modes equidistant in $\zm$ across the last ray, ie. by the remnant. This is illustrated in Figure \ref{fig:EntModes}. (In terms of the affine coordinate $\ym$, in the case equivalent to a constant velocity mirror, $\ym = k \zm$, the Hawking radiation is contained in an interval of length $\ln(k)$ and is purified by modes within a distance $k$ past the last ray. In the case at hand, the length of the purifying region scales as $e^{M^*}$). Moreover, because we restrict our consideration to the Hawking-like region, these modes are still the thermal radiation excited above the vacuum at $\Rscriplus$. Thus we can identify the entanglement of the excited Hawking radiation above the vacuum at $\Rscriplus$ as contributing to the large entanglement entropy of the interval.

\subsubsection{Long-Range Behaviour}

While we have identified the modes that purify the Hawking radiation, this doesn't account for all of the excess entanglement in the interval, nor the change in entropy as we vary the endpoints on large scales. Our calculations in Section \ref{subsec:mirrors} showed that the renormalized entropy falls as we expand $\Sigma$ past the last ray, but it did so on a scale set by the overall length of $\Sigma$. Moreover, we could again re-entangle the interval simply by moving the past boundary sufficiently farther away so our original approximation was once again valid.

Because these large variations are in a regime where both boundaries of the interval are far from the radiating region, the long-distance behaviour of the entanglement entropy is due to modes outside of the radiation region; these modes are locally in their vacuum configuration. It may seem puzzling that excess entropy above the vacuum can be due to modes that are locally in their vacuum state. We now show, though, that the difference in entropy with respect to the vacuum is due to the mismatch in the entanglement of these modes across the radiating region. Moreover, unlike when we looked at the state $\scrimin$, the analysis of the entanglement on $\scriplus$ is insensitive to the UV cutoff.

The separation of entanglement entropy from Hawking-like modes and from long-range entanglement of the vacuum is most clear in the limit where a mirror instantaneously changes from rest to a constant velocity in the original frame. Then on $\Rscriplus$, the state is everywhere in vacuum except at the kink in the mirror trajectory, where there is a mismatch in phase in gluing the Rindler modes of the two regions of vacuum together.

The hatted Rindler modes to the future of the kink are shifted by
\begin{equation}
\widehat{v}'_\omega(\zm) = \frac{1}{\sqrt{2 \omega}} e^{-i \omega \ln( \zm)} e^{i \omega \ln( k)} \Theta(\zm) \, ,
\label{eq:rindler_shift}
\end{equation}
where $\zm = k \zp$ after the kink. In terms of the localized modes described above, this simply induces a shift such that $\widehat{v}'_{jn}$ is peaked near $\zm = \exp(-2 \pi n/\epsilon+\ln(k))$ instead of near  $\zm = \exp(-2 \pi n/\epsilon)$. We see that, relative to the true vacuum, the kinked vacuum has modes prior to the kink entangled with modes localized further to the future.

\begin{figure}
	\centering
		\includegraphics[height=7cm]{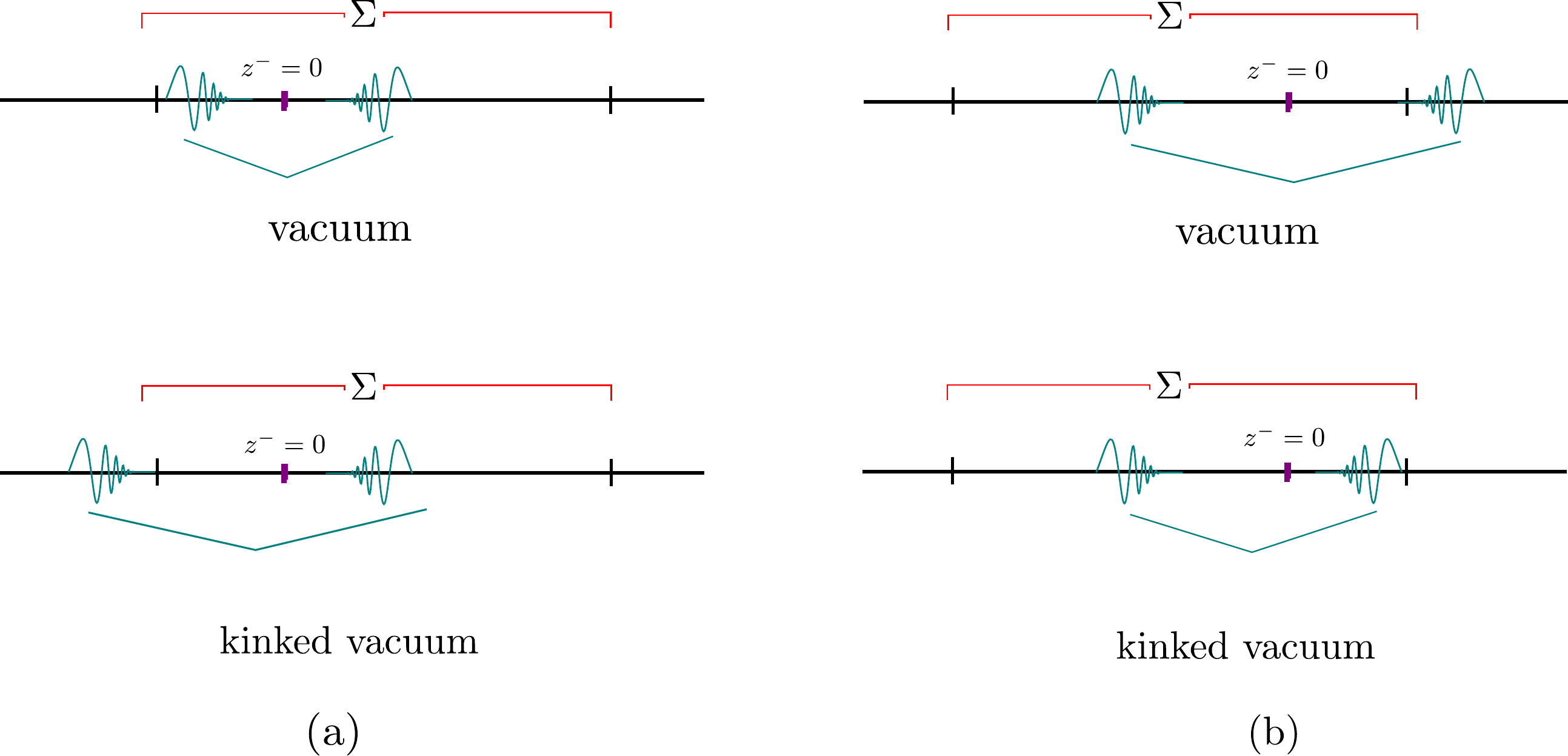}
	\caption{(a) When the interval is much longer to the past of the kink than to the future, there are many modes whose entangled partner is within the future boundary in the vacuum, but outside the boundary in the kinked state. (b) When the interval is much longer to the future of the kink than to the past, there are many modes who entangled partner is outside the past boundary in the vacuum, but inside the boundary in the kinked state.}
	\label{fig:intervals}
\end{figure}

This gives a nice picture of the entanglement entropy due to the mismatched vacuum regions, as illustrated in Figure \ref{fig:intervals}. When the interval is much longer to the future of the kink than it is to the past, there are many modes whose entangled partner across $\zm=0$ is still within the interval in the vacuum state, but has been stretched outside the interval in the kinked vacuum state. This generates the large renormalized entropy.

Conversely, when the interval is much longer to the past of the kink than it is to the future, there are many modes to the future of $\zm=0$ whose entangled partners are outside of the interval in the vacuum state, but are inside of the interval in the kinked vacuum. This generates the negative renormalized entropy found.

%%%%%%%%%%%%%%%%%%%%%%%%%%%%%%%%%%%%%%%%%%%%%%%%%%%%%%%%%%
%%%%%%%%%%%%%%%%%%%%%%%%%%%%%%%%%%%%%%%%%%%%%%%%%%%%%%%%%%
\section{Discussion and Conclusions}\label{sec:discussion}
%%%%%%%%%%%%%%%%%%%%%%%%%%%%%%%%%%%%%%%%%%%%%%%%%%%%%%%%%%
%%%%%%%%%%%%%%%%%%%%%%%%%%%%%%%%%%%%%%%%%%%%%%%%%%%%%%%%%%

%---------------------------------------------------------------%
\subsection{Lifting to Higher Dimensions}\label{sec:lift}
%---------------------------------------------------------------%

It was shown in \cite{Giddings1992} that the CGHS action, \eqref{eq:gravityaction},\eqref{eq:matteraction}, describes the physics of the near horizon region of extremal dilatonic black holes in both four and five dimensions. There are several inequivalent ways of taking the extremal limit which push either the Horizon or the mouth infinitely far away. The classical CGHS action, which gives rise to the black hole solution and the flat dilaton solution, describes the physics of a geometry with a fixed sized throat, and thus corresponds to the limit where the mouth is pushed to infinity. The Horizon limit metric and dilaton are given by,
\begin{align}
ds^2 = {1 \over \kappa^2}\left( -  \tanh^2x \ dt^2  + dx^2 + {d\Omega^2_{(d)} \over (10 - 3 d)} \right), \ e^{2(\phi - \hat{\phi}_0)} =  {\cosh^{-2}x \over   (10 - 3 d) \kappa^2} \label{horizonlimit}
\end{align}
where $d$ is the dimension of the two- or three-sphere on which we reduce to obtain the 2-dimensional CGHS action, and $\hat{\phi}_0$ is the value of the dilaton at infinity, which is held fixed when taking the extremal limit. The overall constant, $\kappa^2$, which appeared as a cosmological constant term in the CGHS action, takes on the values $1/4Q^2$ and $1/Q$ in the four and five dimensional cases respectively, where $Q$ is the magnetic charge associated with a gauge field that points in the $S^{d}$ directions. Taking $x \gg 1$ we obtain the metric and dilaton of the throat limit,
\begin{align}
ds^2 = {1 \over \kappa^2}\left( - dt^2  + dx^2 + {d\Omega^2_{(d)} \over (10 - 3 d)} \right), \ \phi = - x + \hat{\phi}_0 + \ln\left( {2/\kappa \over  \sqrt{10 - 3 d} } \right) \label{throatlimit}
\end{align}
The problem we are studying corresponds to throwing in a pulse into the flat linear dilaton region. In the classical case, the solution corresponds to patching the two solutions (\ref{horizonlimit}) and (\ref{throatlimit}) along the infalling null shockwave (with an appropriate shift of the coordinates). The geometry develops a horizon, which on the null shockwave is located at $x = {1 \over 2} \ln{G M_{ADM} \over \kappa}$. In order for this to carry through to the uplifted case requires $\ln{G M_{ADM} \over \kappa} \gg 1$, since the blackhole in the classical CGHS case forms from throwing a pulse into the flat linear dilaton region. The classical uplifted picture then corresponds to the horizon, of the dilatonic black hole being pulled up the throat by the infalling matter. 

In the mean field approximation scenario of \cite{Ashtekar2011b}, the infalling excitation forms an {\it{apparent}} horizon, which emanates from the null shockwave at  $x \sim {1 \over 2} \ln{G M_{ADM} \over \kappa}$.  To discuss the uplift we have to assume the nature of the geometry past the last ray.  Before the last ray, it was shown that the geometry near $\scri_R^+$ is asymptotically flat and that $dy^-/dz^-$ approaches a constant at the last ray. Assuming that quantum gravity effects are confined to the vicinity of the singularity, we consider the case where the geometry is extended all the way past the last ray near $\scri_R^+$. Thus at late times the geometry transitions back to the throat limit of the extremal dilatonic black hole, although in rescaled coordinates. 

As we have shown, the horizon formed by the infalling excitation evaporates predominantly in the manner of Hawking, producing radiation purely entangled with the region behind the last ray. In the uplifted picture, the radiation is entangled with excitations deep down the throat, and thus the throat geometry itself can be viewed as a remnant. This suggests the same remnant scenario considered previously under the name of {\it{cornucopions}} \cite{Banks:1992ba}. Our conclusion is somewhat unsatisfactory, due to the much-discussed issues with remnants in higher dimensions, including the possibility of infinite remnant pair-production (see, for example, \cite{Banks:1992is, Banks:1992mi,Giddings:1993km, Giddings:1994qt, Susskind:1995da}). Perhaps more importantly, remnants are incompatible with AdS/CFT \cite{Almheiri}. We stress, though, that remnants in the two-dimensional model do not necessarily imply remnants in the uplift, as reduction and quantization do not always commute.\footnote{We thank Don Marolf for emphasizing to us the possibility of markedly different behavior in the uplift.}

%---------------------------------------------------------------%
\subsection{Conclusions}\label{sec:conclusions}
%---------------------------------------------------------------%

We have demonstrated that a black hole in the CGHS model does not result in a firewall, but rather, as previously expected, decays to a highly-entangled remnant: there is a large entanglement between an interval containing the majority of the Hawking flux and the matter state to the future of the last ray. The entanglement scales with the ADM mass of the black hole, while the remnant has a small, universal corrected Bondi mass. 

The lift of the CGHS solution to higher dimensions gives a well-known picture of a near-extremal black hole Hawking radiating back down to extremality. The model then also suggests a higher-dimensional remnant, which lives in the throat outside the extremal horizon. Remnants are problematic due to issues with pair-production and are incompatible with AdS/CFT. While our result may be taken as mildly cautionary to the firewall proposal, we emphasize instead that this is strongly suggestive that the CGHS model, as presently understood, misses essential features of higher-dimensional gravity. Indeed, the model is a renormalizable local field theory and manifestly is not holographic.

It would be much more interesting to study non-local modifications of the CGHS model or altogether different two-dimensional models that better capture the postulates of complementarity. This would be a more useful toy model in which to search for the dynamical (non-)formation of firewalls.

%---------------------------------------------------------------%
%---------------------------------------------------------------%
\section*{Acknowledgements}
\label{sec:acknowledgements}
We thank Sebastian Fischetti, Steve Giddings, Donald Marolf, Eric Mintun, Joseph Polchinski, Mark Srednicki, and Andrew Strominger for useful discussions. This work was supported by NSF grant NSF07-57035.

\bibliographystyle{jhep}
\bibliography{2DBHBib}

\end{document}